\def\NOTES{1}

\documentclass{hotnets22}

\usepackage{times}  
\usepackage[pdfpagelabels=false]{hyperref}
\usepackage{titlesec}
\usepackage{subcaption}
\usepackage{xspace}
\usepackage{enumitem}
\usepackage{xcolor}
\usepackage{todonotes}
\usepackage[locale=US]{siunitx}
\usepackage{tikz}
\usepackage[newfloat=true,frozencache=true,cachedir=.]{minted}
\usepackage{listings}
\usepackage{adjustbox}
\usepackage{url}
\usepackage[T1]{fontenc}

\usepackage{placeins} 
\usepackage{comment}  

\DeclareSIUnit{\nothing}{\relax}

\usepackage{cleveref}

\usepackage[acronym]{glossaries}
\glsdisablehyper

\makeatletter
\newcommand{\DeclareLatinAbbrev}[2]{%
  \DeclareRobustCommand{#1}{%
    \@ifnextchar{.}{\textit{#2}}{%
      \@ifnextchar{,}{\textit{#2.}}{%
        \@ifnextchar{!}{\textit{#2.}}{%
          \@ifnextchar{?}{\textit{#2.}}{%
            \@ifnextchar{)}{\textit{#2.}}{%
              {\textit{#2.,\ }}}}}}}}%
}
\makeatother
\DeclareLatinAbbrev{\eg}{e.g}
\DeclareLatinAbbrev{\Eg}{E.g}
\DeclareLatinAbbrev{\ie}{i.e}
\DeclareLatinAbbrev{\Ie}{I.e}
\DeclareLatinAbbrev{\etc}{etc}
\DeclareLatinAbbrev{\etal}{et~al}

\DeclareDocumentCommand\newstep{o}{%
\item\IfNoValueTF{#1}{}{#1 \textendash\xspace}}

\def\first {$(i)$\xspace}
\def\second{$(ii)$\xspace}

\def\third {$(iii)$\xspace}
\def\fourth{$(iv)$\xspace}
\def\fifth {$(v)$\xspace}

\usepackage{dcolumn}	
\newcolumntype{.}{D{.}{.}{2}}
\DeclareSIUnit\pps{pps}

\newcommand{\systemname}[0]{\textsc{NetBuddy}\xspace}

\newcommand{\kathara}[0]{{Kathará}\xspace}

\newcommand{\smartparagraph}[1]{\vspace{.05in}\noindent\textbf{#1}}

\if \NOTES 1
 \newcommand{\mcnote}[1]{\textcolor{brown}{[marco: #1]}}
 \newcommand{\alnote}[1]{\textcolor{teal}{[alireza: #1]}}
 \newcommand{\chnote}[1]{\textcolor{violet}{[changjie: #1]}}
\else 
 \newcommand{\mcnote}[1]{}
 \newcommand{\alnote}[1]{}
 \newcommand{\chnote}[1]{}
\fi

\newsavebox \FormalSpec
\newsavebox \HighLevelConf
\newsavebox \LowLevelConf

\newcommand*\circled[1]{\tikz[baseline=(char.base)]{
            \node[shape=circle,draw,inner sep=1pt] (char) {#1};}}
            
\newcommand{\circledbf}[1]{\circled{\bf #1}}

\definecolor{SeaGreen}    {RGB}{60,  179, 113}
\definecolor{OliveGreen}  {RGB}{107, 142, 35}
\definecolor{ForestGreen} {RGB}{34,  139, 34}
\definecolor{YellowOrange}{RGB}{255, 204, 0}

\definecolor{DarkBlue}     {rgb}{0.0,  0.0,  0.2}
\definecolor{DarkRed}      {rgb}{0.2,  0.0,  0.0}
\definecolor{HeraldBlue}   {rgb}{0.0,  0.0,  0.8}
\definecolor{HeraldRed}    {rgb}{0.51, 0.12, 0.15}
\definecolor{HeraldRed2}   {rgb}{0.81, 0.12, 0.15}
\definecolor{HeraldGray}   {rgb}{0.2,  0.2,  0.5}
\definecolor{HeraldGreen}  {rgb}{0.0,  0.4,  0.0}
\definecolor{AmberOrange}  {rgb}{1.0,  0.49, 0.0}
\definecolor{CadmiumOrange}{rgb}{0.93, 0.53, 0.18}
\definecolor{ChromeYellow} {rgb}{1.0,  0.65, 0.0}

\newcommand{\refscolor} {blue}
\newcommand{\linkscolor}{HeraldRed2}
\newcommand{\urlscolor} {HeraldBlue}

\hypersetup{
	colorlinks  = true,
	breaklinks  = true,
	linkcolor   = \linkscolor,
	urlcolor    = \urlscolor,
	citecolor   = \refscolor,
	anchorcolor = black
}

\newacronym{ACK}{ACK}{Acknowledgement}
\newacronym{ACL}{ACL}{Access Control List}
\newacronym{API}{API}{Application Programming Interface}
\newacronym{ARP}{ARP}{Address Resolution Protocol}
\newacronym[plural=ASes, firstplural=Autonomous Systems (ASes)] {AS}{AS}{Autonomous System}
\newacronym{ASAP}{ASAP\textsuperscript{2}}{Accelerated Switching and Packet Processing}
\newacronym{BGP}{BGP}{Border Gateway Protocol}
\newacronym{CAT}{CAT}{Cache Allocation Technology}
\newacronym{CDN}{CDN}{Content Delivery Network}
\newacronym{CORD}{CORD}{Central Office Re-architected as a Datacenter}
\newacronym{CPE}{CPE}{Customer Premise Equipment}
\newacronym{CPU}{CPU}{Central Processing Unit}
\newacronym{CSR}{CSR}{Cloud Services Router}
\newacronym{DC}{DC}{datacenter}
\newacronym{DCA}{DCA}{Direct Cache Access}
\newacronym{DDIO}{DDIO}{Data Direct I/O}
\newacronym{DFC}{DFC}{Datapath Flow Cache}
\newacronym{DPU}{DPU}{Data Processing Unit}
\newacronym{FCT}{FCT}{Flow Completion Time}
\newacronym{FIB}{FIB}{Forwarding Information Base}
\newacronym{DUT}{DUT}{Device Under Test}
\newacronym{DPDK}{DPDK}{Data Plane Development Kit}
\newacronym{DPI}{DPI}{Deep Packet Inspection}
\newacronym{DRAM}{DRAM}{Dynamic Random-Access Memory}
\newacronym{DRR}{DRR}{Deficit Round-Robin}
\newacronym{EMC}{EMC}{Exact Match Cache}
\newacronym{ETSI}{ETSI}{European Telecommunications Standards Institute}
\newacronym{FCB}{FCB}{Flow Control Block}
\newacronym{FIFO}{FIFO}{First-In First-Out}
\newacronym{FPGA}{FPGA}{Field-Programmable Gate Array}
\newacronym{FTP}{FTP}{File Transfer Protocol}
\newacronym{GbE}{GbE}{Gigabit Ethernet}
\newacronym{Gbps}{Gbps}{Gigabits per second}
\newacronym{GPT}{GPT}{Generative Pretrained Transformer}
\newacronym{GPU}{GPU}{Graphics Processing Unit}
\newacronym{GRO}{GRO}{Generic Receive Offload}
\newacronym{HBM}{HBM}{High Bandwidth Memory}
\newacronym{HFS}{HFS}{Hadoop Fair Scheduler}
\newacronym{HTTP}{HTTP}{Hypertext Transfer Protocol}
\newacronym{HTTPS}{HTTPS}{Hypertext Transfer Protocol Secure}
\newacronym{IETF}{IETF}{Internet Engineering Task Force}
\newacronym{IDS}{IDS}{Intrusion Detection System}
\newacronym{IO}{I/O}{input-output}
\newacronym{IP}{IP}{Internet Protocol}
\newacronym{IPU}{IPU}{Infrastructure Processing Unit}
\newacronym{IPv4}{IPv4}{Internet Protocol version 4}
\newacronym{IPv6}{IPv6}{Internet Protocol version 6}
\newacronym{ISP}{ISP}{Internet Service Provider}
\newacronym{KVM}{KVM}{Kernel-based Virtual Machine}
\newacronym{LB}{LB}{Load Balancer}
\newacronym{LLC}{LLC}{Last Level Cache}
\newacronym{LLM}{LLM}{Large Language Model}
\newacronym{LRO}{LRO}{Large Receive Offload}
\newacronym{MAC}{MAC}{Medium Access Control}
\newacronym{MPLS}{MPLS}{Multi-Protocol Label Switching}
\newacronym{MTU}{MTU}{Maximum Transmission Unit}
\newacronym{NAPT}{NAPT}{Network Address and Port Translation}
\newacronym{NAT}{NAT}{Network Address Translation}
\newacronym{NETCONF}{NETCONF}{Network Configuration Protocol}
\newacronym{NF}{NF}{Network Function}
\newacronym{NFV}{NFV}{Network Functions Virtualization}
\newacronym{NIC}{NIC}{Network Interface Card}
\newacronym{NLP}{NLP}{Natural Language Processing}
\newacronym{NUMA}{NUMA}{Non-Uniform Memory Access}
\newacronym{OMEC}{OMEC}{Open Mobile Evolved Core}
\newacronym{ONF}{ONF}{Open Networking Foundation}
\newacronym{ONOS}{ONOS}{Open Network Operating System}
\newacronym{OPNFV}{OPNFV}{Open Platform for NFV}
\newacronym{OPP}{OPP}{Open Packet Processor}
\newacronym{OS}{OS}{Operating System}
\newacronym{OVS}{OVS}{Open vSwitch}
\newacronym{P4}{P4}{Programming Protocol-Independent Packet Processors}
\newacronym{PCIe}{PCIe}{Peripheral Component Interconnect Express}
\newacronym{PDN_GW}{PDN GW}{Packet Data Network Gateway}
\newacronym{PIFO}{PIFO}{Push-In-First-Out}
\newacronym{PGW}{PGW}{Packet Gateway}
\newacronym{PMD}{PMD}{Poll Mode Driver}
\newacronym{PoP}{PoP}{Points of Presence}
\newacronym{pps}{pps}{packets per second}
\newacronym{QPI}{QPI}{Quick Path Interconnect}
\newacronym{RAM}{RAM}{Random Access Memory}
\newacronym{RDMA}{RDMA}{Remote Direct Memory Access}
\newacronym{RIB}{RIB}{Routing Information Base}
\newacronym{RLHF}{RLHF}{Reinforcement Learning from Human Feedback}
\newacronym{RMT}{RMT}{Reconfigurable Match Tables}
\newacronym{RR}{RR}{Round-Robin}
\newacronym{RSC}{RSC}{Receive Side Coalescing}
\newacronym{RSS}{RSS}{Receive-Side Scaling}
\newacronym{RQRR}{RQRR}{Resilient Quantum Round-Robin}
\newacronym{SDN}{SDN}{Software-Defined Networking}
\newacronym{SLO}{SLO}{Service Level Objective}
\newacronym{SMC}{SMC}{Signature Match Cache}
\newacronym{SMT}{SMT}{Satisfiability Modulo Theories}
\newacronym{SNF}{SNF}{Synthesized Network Functions}
\newacronym{SNMP}{SNMP}{Simple Network Management Protocol}
\newacronym{SRAM}{SRAM}{Static Random Access Memory}
\newacronym{SRIOV}{SR-IOV}{Single Root I/O Virtualization}
\newacronym{TBF}{TBF}{Token Bucket Filter}
\newacronym{TCAM}{TCAM}{Ternary Content-Addressable Memory}
\newacronym{TCP}{TCP}{Transmission Control Protocol}
\newacronym{TSS}{TSS}{Tuple-Space Search}
\newacronym{TTL}{TTL}{Time to Live}
\newacronym{UDP}{UDP}{User Datagram Protocol}
\newacronym{VLAN}{VLAN}{Virtual Local Area Network}
\newacronym{VM}{VM}{Virtual Machine}
\newacronym{VMDq}{VMDq}{Virtual Machine Device queues}
\newacronym{VNIC}{VNIC}{Virtual NIC}
\newacronym{VPN}{VPN}{Virtual Private Network}

\hypersetup{pdfstartview=FitH,pdfpagelayout=SinglePage}

\begin{document}


\title{Making Network Configuration Human Friendly}
\author{Changjie Wang$^\dag$, Mariano Scazzariello$^\ddag$, Alireza Farshin$^*$, Dejan Kosti\'c$^\ddag$, Marco Chiesa$^\ddag$\\
$^\dag$ École Polytechnique $^\ddag$ KTH Royal Institute of Technology $^*$ RISE Research Institutes of Sweden}



\maketitle

\begin{abstract}

This paper explores opportunities to utilize Large Language Models (LLMs) to make network configuration human-friendly, simplifying the configuration of network devices and minimizing errors. We examine the effectiveness of these models in translating high-level policies and requirements (\ie specified in natural language) into low-level network APIs, which requires understanding the hardware and protocols. More specifically, we propose \systemname for generating network configurations from scratch \textit{and} modifying them at runtime. \systemname splits the generation of network configurations into fine-grained steps and relies on self-healing code-generation approaches to better take advantage of the full potential of LLMs. We first thoroughly examine the challenges of using these models to produce a fully functional \& correct configuration, and then evaluate the feasibility of realizing \systemname by building a proof-of-concept solution using GPT-4 to translate a set of high-level requirements into P4 and BGP configurations and run them using the \kathara network emulator. 

\end{abstract}

\section{Introduction}
\label{sec:introduction}

Networks are the backbone of today's communication infrastructure, powering everything from simple online interactions to mission-critical services. Network operators wield significant control over the flow of data in a network, guiding it along its journey from one device to the next by carefully specifying a set of per-device configurations within the network infrastructure. These configurations -- which can affect devices \& services ranging from switches/routers, servers, network interfaces
, network functions
, and even GPU clusters (
used for training or inference of online services) -- must be carefully configured to ensure the reliable transmission of information.

In the last decade, academia and industry adopted \gls{SDN} to simplify the configuration of networks 
compared to the previous traditional (monolithic) paradigm. 
Despite the benefits brought by \gls{SDN}, network configuration entails frequent human intervention.
Manual configuration is costly (\eg requires expert developers who know each of the different APIs and protocols), difficult, and susceptible to human error; the consequence of errors can be dire (\eg a 911 emergency call outage
\,\cite{911_failure}, a disruption to flight schedules at Amadeus
\,\cite{airport_outage}, and a 
outage at Meta
\,\cite{meta_outage}).


Many efforts have attempted to simplify the process of compiling a high-level policy specified by a network operator into a set of per-device network configurations\,\cite{propane_at, synet, netcomplete, jinjing, contra, aura} and to minimize errors by generating configurations with provable guarantees via verification\,\cite{batfish, minesweeper, tiramisu, arc, plankton, shapeshifter, p4v}. However, network configuration remains an arduous, complex, and expensive task for network operators. Moreover, the abstraction and composition of networks force these configuration tools to employ a self\nobreakdash-defined specification or language for succinctly describing network intents. For instance, SyNet\,\cite{synet} introduced a stratified Datalog to express routing protocols and network requirements. These approaches impose another challenge on network operators, they must acquire proficiency in a new domain-specific language that may not be widely used and could potentially have flaws.

\smartparagraph{What has changed?} Generative AI and \glspl{LLM} (\eg OpenAI's GPT\,\cite{gpt_3_paper,gpt_4_paper}, Google's PaLM\,\cite{google_palm,google_palm2}, and Meta's LLaMA\,\cite{meta_llama}) have recently shown great potential to generate coherent, contextually relevant content, answer questions, and even engage in meaningful conversations with users; all of these capabilities offer immense potential for applications in various industries. For instance, GitHub Copilot, Amazon CodeWhisperer, StarCoder\,\cite{starcoder}, and WizardCoder\,\cite{wizardcoder} offer services using \glspl{LLM} to perform coding tasks. Additionally, some systems (\eg AutoGPT\,\cite{autogpt}) try to solve large problems. 

In this paper, we plan to answer a scientific question in our domain of whether \glspl{LLM} can improve network configuration. \glspl{LLM} enable new possibilities in \textit{quickly} acquiring \textit{vast} knowledge (\eg they can 
learn all IETF documents, protocol specifications, standards, and best practices), which goes beyond the capabilities of humans and state-of-the-art works on network configuration. We explore various opportunities to simplify and potentially automate the configuration of network devices based on human language prompts/inputs, closing the gap between network operators and network control with the help of AI. 
Using AI for networking has already been explored by previous works, such as Ben-Houdi \etal\,\cite{nlp_configuration} that  presented potential use cases for taking advantage of \gls{NLP} techniques for networking. Additionally, Nile\,\cite{intent_based_synthesis} introduced an intent-based networking scheme where \gls{NLP} is applied in translation to a simple and limited set of requirements expressed in basic natural language.

In this paper, we go two steps further by \first embarking upon the systematic study of utilizing emerging \glspl{LLM} to configure network devices from high-level requirements specified in natural language, and \second showing a \textit{functional} LLM-based prototype and discuss the related challenges faced while building such a prototype. Our main results show that state-of-the-art \glspl{LLM} are capable of generating fully working P4 \& BGP configurations (without fine-tuning) to enforce a class of path policies expressed as natural language requirements. We show that there exists an inherent trade-off between the accuracy of an LLM model, the complexity of the requirements, and the economic cost. 


Although we target switch/router configuration, the same principle could apply to other network configurations (see 
\S\ref{fig:design:netbuddy}) and tasks. For instance, \glspl{LLM} can potentially simplify the cumbersome management of Kubernetes clusters or be used for network troubleshooting. 
 We expect future networks to increasingly rely on AI techniques to perform many tasks. Some of these tasks may benefit from \glspl{LLM}, whereas others may require different techniques. 
Our work is still underway, but we hope it enables further research on effectively using AI techniques to address the challenges of network management/configuration.

\smartparagraph{Contributions.} In this paper, we: 
\begin{itemize}[nosep,leftmargin=*]
    \item Propose a system called \systemname to simplify \& potentially automate the configuration of network devices; 
    \item Examine the challenges 
    of using \glspl{LLM} for network configuration and evaluate the effectiveness of GPT-4 in translating input requirements to a formal specification;
    \item Present proof-of-concept solutions to demonstrate the feasibility of \systemname; 
    it can generate P4 configurations from scratch and adapt existing BGP configurations.
\end{itemize}

\section{From High-level Requirements\\ to Low-Level Network Configurations}
\label{sec:idea}

\begin{figure}[th!]
    \centering
    \includegraphics[width=0.96\columnwidth,page=1, trim = 2mm 0mm 2mm 52mm, clip]{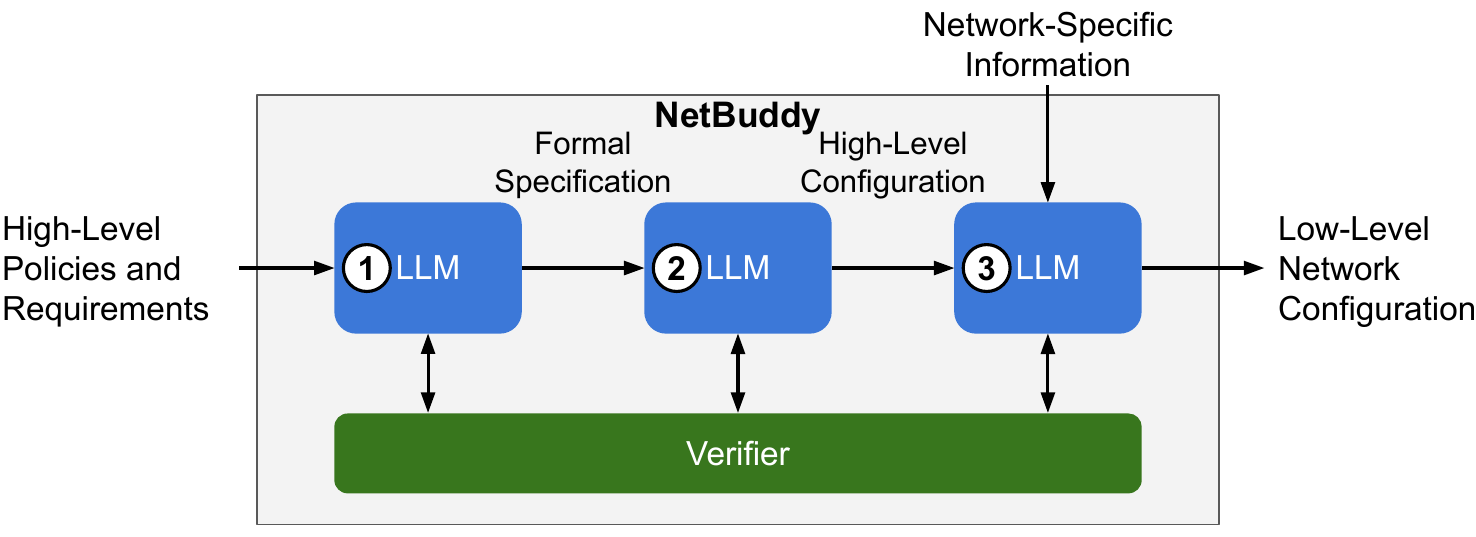}
    \caption{High-level overview of \systemname, where the high-level policies \& requirements are translated in low-level network configuration in three steps.
    }
    \label{fig:design:netbuddy}
    \vspace{-1em}
\end{figure}

This section presents \systemname, our proposed LLM-based network configuration generator. \systemname receives high-level policies and requirements (\ie specified in natural language) from network operators and translates them into low-level network configurations (\eg API calls \& commands). Since \textit{directly translating} natural language to low-level network configurations may be too complex for the existing \glspl{LLM}, and to facilitate the understandability of what the reasoning for the configuration is, we propose a multi-stage pipeline for performing such a translation in multiple less-complex \& fine-grained steps (see Fig.\,\ref{fig:design:netbuddy}). Our design is inspired by recently published works that try to improve the reasoning \& performance of \glspl{LLM} by prompt engineering (\eg chain-of-thought prompting\,\cite{cot_prompting}) and using self-healing\,\cite{selfhealing_code,compile_feedback} \& self-collaborative code generation\,\cite{self_collaborative_code}.

\systemname translates the high-level policies \& requirements to low-level network configurations in three steps, each may be performed by an \gls{LLM} (see Fig.\,\ref{fig:design:netbuddy}).  Network operators may adapt \systemname to their protocols and use cases to \first operate with a different number of steps and \second integrate other components \& tools. In this work, we focus on general-purpose \glspl{LLM}, but a network operator can potentially use different (expert/specialized) \glspl{LLM} for each task, \ie pre-trained or fine-tuned for a specific domain. The current trend is to fine-tune multi-billion-parameter \glspl{LLM} since the costs are rather low and the resources needed for inferencing are much less than those required for a general-purpose \gls{LLM} with multi-trillion parameters; however, a general-purpose \gls{LLM} that is capable of correctly performing the tasks may be preferred over an expert/specialized \gls{LLM} due to higher resource utilization (similar to the popularity of general-purpose processors over specialized hardware/accelerators). In addition to \glspl{LLM}, \systemname also utilizes a verifier component to verify the output of each \gls{LLM} and to provide feedback in case it detects errors/flaws, which is an essential component for a fully automated pipeline. We assume that network operators already have the machinery to deploy and/or run the generated low-level configuration on the network devices; therefore, we do not consider this step. 

\smartparagraph{\circled{1} Generating formal specification.} Initially, \systemname translates the received natural language input to a \textit{pre\nobreakdash-defined} formal specification format (\eg a data structure or a domain-specific language). The goal of this step is to convert the network operators' input into a known format that is easier to verify; this is challenging since we need to ensure the correctness and stability of the translations (see \S\ref{sec:challenges}). For instance, the input information can be converted to a simple data structure to specify the reachability, waypoints, and avoidances in a network, as these are simple requirements heavily used by the state-of-the-art\,\cite{netkat, genesis, synet, config2spec, ryanthesis}. Depending on the complexity of the network requirements \& policies, a network operator may directly add/remove new entries in the formal specification format (\eg to consider link preferences and/or load balancing to more efficiently route traffic). \Cref{lst:design:formal_specification} shows a sample input and its formal specification; the user provides the sample input and the format of the data structure.

\begin{lrbox}{\FormalSpec}\begin{minipage}{\columnwidth}
\begin{minted}[frame=single,framesep=8pt,fontsize=\footnotesize,tabsize=4,framerule=0pt]{json}
{
  "reachability": {
    "s1": ["h1", "h2"],
    "s2": ["h1", "h2"],
    "s3": ["h1", "h2"],
    "s4": ["h1", "h2"],
  },
  "waypoint": {
    ["s1", "h1"]: ["s2"],
  },
  "avoidance": {
    ["s4", "h2"]: ["s3"],
  }
}
\end{minted}
\end{minipage}\end{lrbox}

\begin{listing}[ht!]
\centering
\caption{An example data structure (bottom
) representing a formal specification for a sample input (top
). \emph{Reachability} refers to the ability of traffic originating from a switch to reach a destination host. \emph{Waypoint} forces the traffic from a switch towards a host to traverse a predefined sequence of intermediate switches. \emph{Avoidance}, in contrast, prevents the traffic traversing certain switches.}
\label{lst:design:formal_specification}

\noindent\begin{minipage}{0.73\columnwidth}
\small
\textbf{Network components:}
\begin{itemize}[nosep,leftmargin=*]
    \item 4 switches: \texttt{s1}, \texttt{s2}, \texttt{s3}, \texttt{s4}
    \item 2 end-hosts: \texttt{h1}, \texttt{h2} 
\end{itemize}

\textbf{Requirements:}
\begin{itemize}[nosep,leftmargin=*]
    \item All the switches can reach all the destination hosts.
    \item Traffic from \texttt{s1} to \texttt{h1} should travel across \texttt{s2}.
    \item To reach \texttt{h2}, \texttt{s4} needs to avoid \texttt{s3}.
\end{itemize}
\noindent\rule{\columnwidth}{0.5pt}
\end{minipage}
\resizebox{0.75\columnwidth}{!}{\usebox\FormalSpec}
\vspace{-2em}
\end{listing}

We mainly assume that the input is provided in natural language format, and it contains a high-level description of the network along with new policies \& requirements. However, by using a multi-modal \gls{LLM}, it would be possible to include other information (\eg notifications/emails from a network monitoring service) in other formats (\eg the image/graph of the network topology). Furthermore, since there is currently no technique to evaluate the correctness of the translation from the natural language input, the verifier component can only check the format and syntax of the generated output or conflicts\slash bad-practices in the requirements; the correctness of the output should be checked by the network operator (see \S\ref{sec:challenges} for more details). 

\smartparagraph{\circled{2} Generating high-level configurations.} Before generating low-level network configurations, \systemname translates the formal specification into a high-level configuration.\footnote{This step also requires the physical topology; therefore, this topology has to be provided as an input if it was not considered
.
} By introducing this intermediate state, \systemname simplifies the task for the \gls{LLM} and facilitates validation of the generated output (\eg \gls{RIB} or \gls{FIB}). \Cref{lst:design:high_level_conf} shows an example of the routing information for 
\Cref{lst:design:formal_specification}.

Since \glspl{LLM} are incapable of computation and execution, prompting \glspl{LLM} to develop code to generate the routing information often results in more efficient \& correct output (see \S\ref{sec:challenges} for more details). In this optional step, the verifier component can be used to \first check the syntax, compilability/runnability, and correctness of the generated output (script and/or configuration) and \second provide feedback to the \gls{LLM} to improve the output. We do not generate the code 
unless it is needed, as this is an expensive step. For instance, if an operator modifies a requirement (\eg remove connectivity between two nodes) but does not modify the type of requirements (\eg suddenly deciding to add traffic priorities), then we re-use the already generated code and only provide the new formal specification as input to the code. More precisely, as long as \textit{no new} types of requirements (\eg traffic priorities) are added, we do not rely on the \gls{LLM} to generate any new code. 
In case an operator adds a new type of requirement (\eg traffic priorities), an LLM would receive the new requirements along with the previous versions of the code producing high-level configurations, including the verifier program and some test cases, to extend the previous programs accordingly to the new requirements \textit{rather than} developing them from scratch.

\begin{lrbox}{\HighLevelConf}\begin{minipage}{\columnwidth}
\begin{minted}[frame=single,framesep=2pt,framerule=0pt,fontsize=\large,tabsize=4]{json}
{
  "h1": {
    "h2": ["s4", "s2", "s1"], 
  }, 
  "h2": {
    "h1": ["s1", "s2", "s4"], 
  }, 
}
\end{minted}
\end{minipage}\end{lrbox}

\begin{listing}[t!]
\centering
\caption{An example topology (left
) and routing information acting as the high-level configuration (right
). This data structure shows the shortest path between every pair of end-hosts.}
\label{lst:design:high_level_conf}

\noindent
\begin{minipage}[!b]{0.45\columnwidth}
\centering
\resizebox{1\columnwidth}{!}{
\centering
\begin{tikzpicture}[thick]
  \node (h2) {\texttt{h2}};
  \node[right=of h2] (s1) {\texttt{s1}};
  \node[above right=of s1] (s2) {\texttt{s2}};
  \node[below right=of s1] (s3) {\texttt{s3}};
  \node[below right=of s2] (s4) {\texttt{s4}};
  \node[right=of s4] (h1) {\texttt{h1}};
  
  \path[<->] (h2) edge (s1)
             (s1) edge (s2)
             (s1) edge (s3)
             (s2) edge (s3)
             (s2) edge (s4)
             (s3) edge (s4)
             (h1) edge (s4);
\end{tikzpicture}}
\end{minipage}
\vrule
\begin{minipage}[!b]{0.478\columnwidth}
\resizebox{1\columnwidth}{!}{\usebox\HighLevelConf}
\end{minipage}
\vspace{-1.5em}
\end{listing}

\smartparagraph{\circled{3} Generating low-level configurations.} As the final step, \systemname generates low-level network configurations, which is very challenging, as we have to deal with a plethora of networking protocols and various vendor-specific syntaxes. Since network configuration is done at different levels and scales, from configuring network interfaces (\eg specifying IP addresses) to routing paths on the Internet (\eg setting up BGP and MPLS), the output of this step may greatly differ in different networks and use cases. Some essential network configurations (along with example tools, APIs, and protocols) are: \first host/server configurations (\eg 
\texttt{iproute}, DHCP, and DNS), \second virtual machines and container configurations (\eg OpenStack, 
Docker, and Kubernetes), \third NIC, IPU, and DPU (\eg P4, eBPF, and Verilog/VHDL), \fourth switch/router configurations (\eg OpenFlow, P4, MPLS, and BGP), and \fifth network function configurations (\eg Click, BESS, and VPP). In this paper, we mainly focus on the fourth group and assess the effectiveness of \systemname in generating P4 and BGP configurations. Generating low-level configurations may require additional information about the underlying devices (\eg MAC and IP addresses of different interfaces), and this information should be provided as input to the \gls{LLM}. Similar to Step \circledbf{2}, the output of this step could be a script to generate a low-level configuration based on the pre-defined routing information, and the verifier component can check the syntax, compilability/runnability, and correctness of the generated output 
and provide feedback if necessary. For instance, one can use existing verification tools (\eg SwitchV\,\cite{switchv} to validate P4 control plane configurations, or Minesweeper\,\cite{minesweeper} to check BGP configurations) to ensure the correctness.

In the majority of the cases, the configuration, commands, and protocol specification may be known by a general\nobreakdash-purpose \gls{LLM} to generate the low-level network configuration.  However, there are cases where a network uses proprietary devices and/or protocols, the details of which are not included in \glspl{LLM}. In such cases, \systemname needs to receive additional \textit{``Network-Specific Information''} (\eg protocol standards, API \& commands documentation, and device characteristics \& performance models) to fine-tune an existing \gls{LLM}. The network documentation and protocol standards are often lengthy, which makes the network configuration by (human) developers time\nobreakdash-consuming and expensive. In contrast, \glspl{LLM} can be fine-tuned in a few hours/days\,\cite{llama_adapter_1,llama_adapter_2,lit_llama} (depending on the amount of additional information) and configure the network more efficiently and cheaper than humans. The average annual salary of a network engineer could be around 100k USD\,\cite{network_eng_salary}, whereas inferencing via GPT-4 would probably cost less than a few thousand USD\,\cite{gpt_4_price,gpt_4_price2}.

\section{Challenges and Takeaways}
\label{sec:challenges}

This section discusses the potential challenges of using \glspl{LLM} to build \systemname. We discuss each challenge 
based on our experience 
with GPT-4.

\smartparagraph{How to ensure correctness when converting natural language to a formal specification?} 
Currently, there is no technique to evaluate the correctness of the translation from natural language to formal specification. However, one can increase the confidence in the result with a suitable dataset that can be used to fine-tune \glspl{LLM} or utilize techniques for evaluating the quality of the translation (\eg via \gls{RLHF}\,\cite{openai_rlhf}). To better understand the effectiveness of existing \glspl{LLM}, we 
evaluated the correctness of the translation in various scenarios when using GPT-4. Nevertheless, providing a definitive evaluation is challenging due to the instability of \glspl{LLM}\,\cite{hallucinations_LLM} and the ambiguity of natural languages.

\smartparagraph{\circled{1} Contradictory requirements.} When multiple entities (\eg customers, managers, operators) specify their requirements individually, their combination may contain contradictory requirements. Even when a single network operator is involved, contradictory network requirements can still occur, especially when the number of conditions is large. These conflicts can result in serious configuration failures before operators discover them, but previous network synthesizers have often overlooked this aspect. We manually crafted some requirements with explicit conflicts and observed that GPT-4 is capable of detecting them (see \emph{Explicit Requirements} in \Cref{lst:challenges:conflicts}). In contrast, identifying \textit{implicit} conflicts between requirements can be more challenging; hence, without fine-tuning, \glspl{LLM} may struggle to directly detect these contradictions. For instance, as shown in \Cref{lst:challenges:conflicts}, we have \emph{Implicit Requirements} where \texttt{s1} needs to reach \texttt{h2} through \texttt{s2} but  \texttt{s2} itself cannot reach \texttt{h2}. GPT-4 translates to \emph{Reachability} and \emph{Waypoint} specification \textit{without} reporting any conflicts. In contrast, it is possible to use SMT solvers \cite{smt_conflict} to ensure there exists \textit{a solution} for a given formal specification. In case of detecting any contradictions, an \gls{LLM} can interpret them and provide feedback to network operators. 
    
\begin{listing}[t!]
\centering
\caption{GPT-4 is capable of detecting explicit contradictions in the input requirements.}
\label{lst:challenges:conflicts}

\noindent\begin{minipage}{0.74\columnwidth}
\small
\textbf{Explicit Requirements:}
\begin{itemize}[nosep,leftmargin=*]
    \item \texttt{s1} can reach \texttt{h1} and \texttt{h2}, but not \texttt{h3}.
    \item \texttt{s1} cannot send traffic to \texttt{h2}.
\end{itemize}
\textbf{GPT-4's Output:}
It's not possible for \texttt{s1} to reach \texttt{h2} and at the same time for \texttt{s1} not to send traffic to \texttt{h2}.
\end{minipage}\vspace{0.2em}
\noindent\begin{minipage}{0.74\columnwidth}
\small
\noindent\rule{\columnwidth}{0.5pt}

\textbf{Implicit Requirements:}
\begin{itemize}[nosep,leftmargin=*]
    \item \texttt{s2} cannot reach \texttt{h2}, \texttt{s1} can reach \texttt{h2}.
    \item \texttt{s1} should forward traffic via \texttt{s2} to reach \texttt{h2}.
\end{itemize}
\end{minipage}
\vspace{-1em}
\end{listing}

\begin{figure}[tbh!]
    \centering
    \includegraphics[width=0.9\columnwidth]{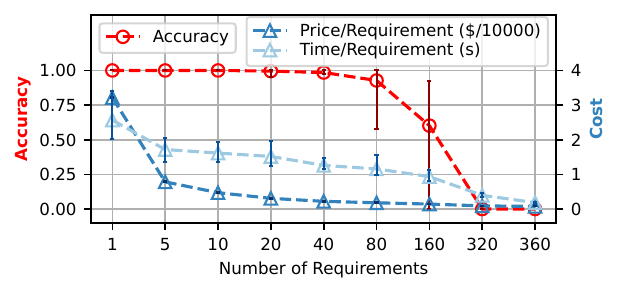}
    \caption{GPT-4 achieves the highest accuracy when translating batches containing 10 requirements. The left y-axis (red color) shows the accuracy for different numbers of requirements; the right y-axis (blue color) shows the per-requirement costs when using GPT-4 API. The temperature parameter is set to 0. Each point shows the median of 10 runs with min/max error bars.}
    \label{fig:challenge:accuracy_cost}
    \vspace{-1.5em}
\end{figure}

\smartparagraph{\circled{2} Complexity.} To understand whether existing \glspl{LLM} are sufficient for our purpose, we devised an experiment as follows: we \first generated 802 network requirements focusing on reachability and waypoints using Config2Spec\,\cite{config2spec}; \second converted them into formal specification format; \third translated them to natural language using GPT-4 based on predefined templates; and \fourth evaluated the efficiency of GPT-4 by comparing the translated version of formal specification with their original format. Fig.\,\ref{fig:challenge:accuracy_cost} shows the accuracy and costs (\ie per-requirement time \& price) of using GPT-4 API\footnote{We observed that the results are more stable when the temperature parameter is set to 0-0.2, as the model is more deterministic\,\cite{gpt_4_temperature}.} for different numbers of input requirements. GPT-4 can \textit{correctly} output formal specifications when the number of input requirements is below 40 per prompt. In contrast, GPT-4's output becomes unstable for higher numbers of requirements, especially when getting closer to GPT-4 input tokens limit (\ie 8192): \first it cannot remember and map all the inputs
; \second the API fails to generate a response (such errors explain the wide error bars depicted in Fig.\,\ref{fig:challenge:accuracy_cost}); and \third it returns truncated responses, causing errors during parsing (more evident 
with 320 and 360 requirements).

Our results, however, also suggest that increasing the number of requirements reduces per-requirement synthesis cost. For instance, translating 10 requirements in one message is  $\sim6\times$ cheaper than translating 10  requirements with 10 messages while still achieving 100\% accuracy. Therefore, batching of requirements  ensures cost-effectiveness and guarantees correctness. Our experiments prove that 360 requirements can be correctly translated by dividing them into 36 batches of 10 requirements.

\smartparagraph{\circled{3} Ambiguity.} The ambiguity of human language and unfamiliarity with specific classes of problems may result in misinterpretations. To better understand this case, we \first used GPT-4 to re-generate some requirements with varying ambiguity levels (defined by GPT-4 itself) and then \second used them to generate a formal specification. We observed that GPT-4 can successfully translate the input requirements, as long as they are not too general and unclear. In cases with higher levels of ambiguity, 
GPT is more prone to make mistakes. For instance, GPT-4 converted a clear requirement as ``\textit{Switch \texttt{s1} should have direct reachability to hosts \texttt{h1}, \texttt{h2}, and \texttt{h3}}'' to a hard-to-interpret ``\textit{Facilitate the exploration of potential communication possibilities between \texttt{s1} and any of the end-hosts (\texttt{h1}, \texttt{h2}, and \texttt{h3})}'', which may be misleading even for a human to understand.

\smartparagraph{Can the output of an \gls{LLM} be immediately passed to the next step?} \glspl{LLM} often provide additional explanations in their output, even when explicitly asked not to give any. Domain-specific languages for \glspl{LLM} (\eg LMQL\,\cite{lmql}) can potentially facilitate the interaction with these models and address this issue. Asking GPT-4 to act like an API server and to put the output configuration/code in a JSON structure with a particular tag simplifies the parsing of the output, making it possible to immediately feed the output to another step. Additionally, using the function calling\,\cite{function_calling_api} feature can alleviate this problem, by which each input requirement can be translated into a function call that modifies the formal specification (\eg adds an entry). A function call is expected to produce a user-defined JSON format, but it is prone to errors (as also observed by us). GPT-4 currently supports only one function call per message, which cannot satisfy advanced requirements that require multiple function calls.

\smartparagraph{What is the best way of using an \gls{LLM} for the task at hand?} 
Prompting \glspl{LLM} to develop a script/code for a problem often results in better output than asking directly for a response (as opposed to\,\cite{bgp_blog_post}). For example, our experiments show that GPT-4 fails to \textit{directly} translate formal specification to the routing information, whereas it can generate a functional Python code to perform the task. We noticed that providing detailed instructions for writing the code can greatly improve code generation and minimize potential errors. For instance, one can use the following instructions to develop a Python script for calculating the shortest path between different end-hosts: ``\first construct a graph from the topology; \second identify the unidirectional host pairs based on the network topology; \third find all possible paths for each host pair, and rank the paths according to their length; \fourth pick the path that strictly satisfies all the requirements related to the switches - if no path is found, set the path to $[]$; and \fifth return final routing paths for all host pairs.''. Such an instruction can be produced by the same or another (specialized) \gls{LLM}
, and later it can be input together with other information to develop the code. Finally, the generated output script/code may contain syntax and functional errors. To solve this issue, \systemname also embeds a \emph{Verifier} component (see Fig.\,\ref{fig:design:netbuddy}) that can detect such issues (via existing tools, real or symbolic execution engines, and pre-defined test cases) and provide feedback to the \gls{LLM} iteratively until all the syntax and functional errors are fixed.

\section{Proof-of-Concept Utilizing GPT-4}
\label{sec:poc}

This section demonstrates the feasibility of realizing \systemname by showcasing two examples of network configuration using GPT-4. 
We will release our code and GPT prompts. In our current prototype, we verify outputs and provide feedback manually and leave the implementation of an automated verifier as future work.

\subsection{MPLS Routing for P4-Enabled Switches}
\label{subsec:mpls}

\begin{figure}[htb!]
    \centering
    \subfloat[\small The network topology.]{
        \includegraphics[width=0.9\columnwidth,,page=3, trim = 5mm 30mm 5mm 20mm, clip]{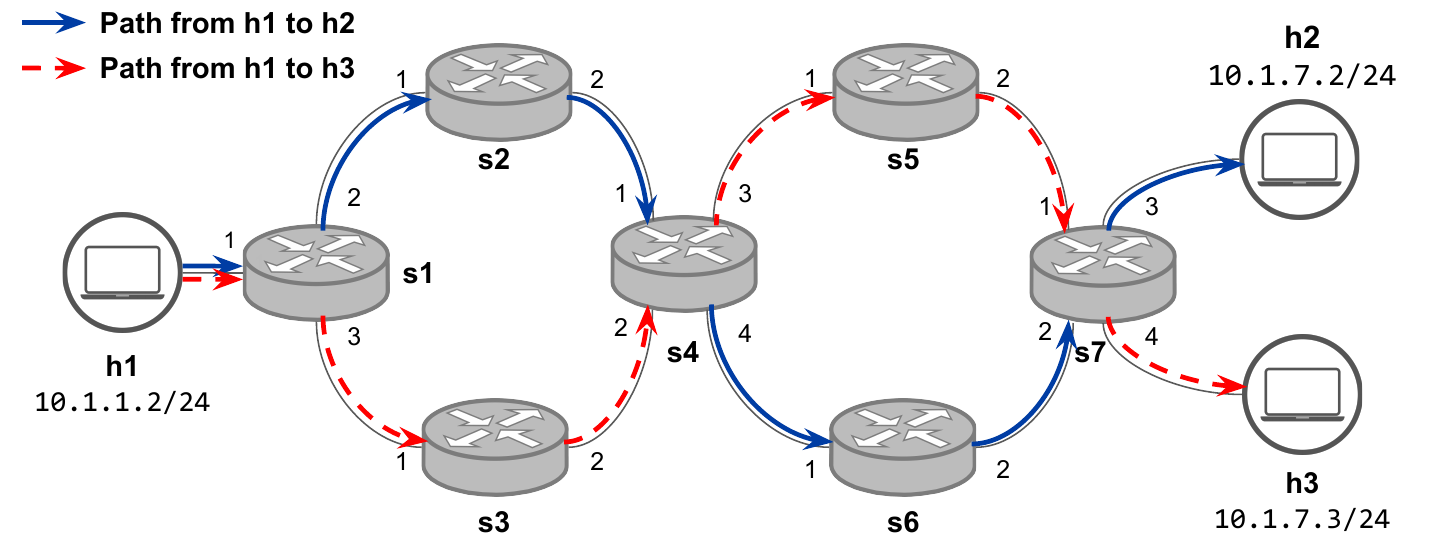}\label{fig:poc:mpls:topology}
    }
    
    \subfloat[\small The input/output of \systemname.]{
        \includegraphics[width=0.9\columnwidth,page=4, trim = 2mm 25mm 8mm 20mm, clip]{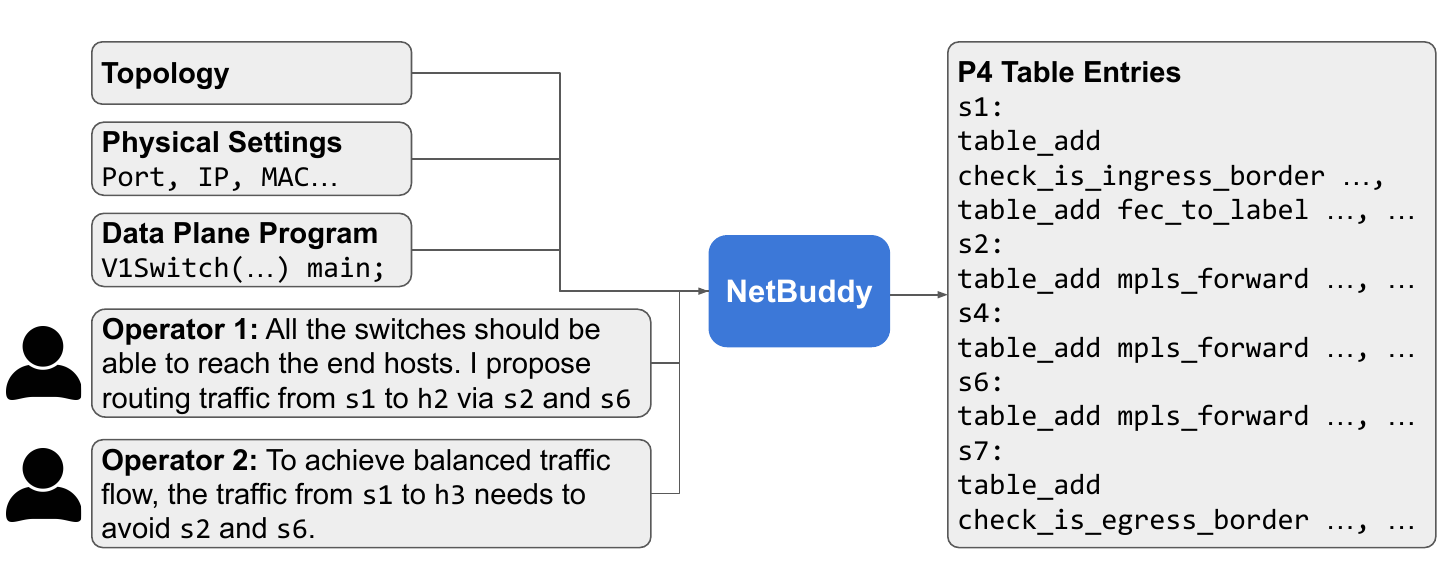}\label{fig:poc:mpls:system}
    }
    \caption{\systemname takes inputs from network operators and generates P4 table entries to route traffic according to the input requirements. The red and blue links in the top figure show the different paths for the traffic between end-hosts.}
    \label{fig:poc:mpls}
    \vspace{-1em}
\end{figure}

In the first scenario, we aim to demonstrate the feasibility of generating network configuration from scratch using \systemname. In particular, we focus on MPLS routing via P4-enabled switches. \systemname receives two sets of information as input: \first the essential details of the network (\ie the topology, physical settings, and data plane program) and \second the network requirements in natural language (see Fig.\,\ref{fig:poc:mpls:system}), and generates P4 table entries for each switch such that the entries satisfy the input requirements
. 

Fig.\,\ref{fig:poc:mpls:topology} shows the evaluated network topology, where three end-hosts (\texttt{h1}-
\texttt{h3}) communicate with each other via seven switches (\ie \texttt{s1}-\texttt{s7}). We assume that the P4 data plane program is already provided,\footnote{GPT-4 is currently unable to generate entire P4 programs, but this may be addressed by fine-tuning.} and network devices are configured with MAC and IP addresses. Next, we explain the details of each step to generate P4 table entries (according to \S\ref{sec:idea}). Fig.\,\ref{fig:poc:mpls:system} shows various inputs and the corresponding outputs provided to and generated by \systemname.

\smartparagraph{Step \circled{1}} Initially, we assume \texttt{Operator 1} 
states two requirements: \first full reachability and \second traffic engineering via waypoints. \systemname converts them into formal specification with a format that includes reachability and waypoint, similar to that in \Cref{lst:design:formal_specification}. 


\smartparagraph{Step \circled{2}} \systemname receives the output of the previous step 
and the network topology to generate forwarding paths. We use the same data structure presented in \Cref{lst:design:high_level_conf} for the high-level network configuration to specify the routing information. To do so, we ask \systemname to generate a Python script to calculate the forwarding paths from the input information. \systemname interacts with GPT-4 at least two times to ensure its understanding of \first the input and output format; \second the desired functionality. These interactions may contain feedback to fix potential errors. In this example, the output of the script contains six forwarding paths. The path from \texttt{h1} to \texttt{h2} (blue) 
satisfies the requirement of waypoint.


\smartparagraph{Step \circled{3}} \systemname interacts with GPT-4 three times to sequentially provide \first the topology, \second switches \& hosts configurations (\eg MAC and IP addresses), and \third the already-deployed P4 program. Then, it uses the output of the previous step 
to generate the P4 table entries. 

In this example, \systemname generates P4 table entries to configure switches for all the forwarding paths between hosts. For instance, for the blue path, the final P4 program performs the following tasks:  \texttt{s1} checks whether the incoming packet originates from \texttt{h1}; if so, it adds an MPLS header to the packet and forwards it to \texttt{s2}. Then, \texttt{s2}, \texttt{s4}, \texttt{s6}, and \texttt{s7} perform MPLS forwarding based on the attached label. Finally, \texttt{s7} removes the MPLS label and transmits the packet to \texttt{h2}. To evaluate the correctness \& runnability of the generated configurations, we emulate the same network using the \kathara network emulator\,\cite{kathara}. Our testbed uses the output of \systemname (\ie a JSON structure) and automatically installs the generated table rules on the P4-enabled switches running on \kathara. Later, we manually test the requirements via \texttt{ping} and \texttt{tcpdump}.

\smartparagraph{Modifying the pipeline to introduce a new requirement.} All forwarding paths, except the blue one, are selected simply based on the shortest path policy. We assume \texttt{Operator 2} introduces a new requirement to improve link utilization and distribute the load more efficiently. To support the new requirement category, the operator adds a new entry called ``avoidance'' (see \Cref{lst:design:formal_specification}) to the specification format. Instead of generating everything from scratch, \systemname can re-use the already-existing configurations to enforce the new requirement. In particular, in Step \circled{2}, \systemname modifies the Python script by interacting with GPT-4 to provide \first the existing script, \second the explanation of the script, \third the details of the new requirement (\ie avoidance) \& updated specification format, and \fourth additional feedback in case of error. The new script calculates new forwarding paths based on the newly-introduced requirement, which forces the traffic from \texttt{h1} to \texttt{h3} to go through the red path shown in Fig.\,\ref{fig:poc:mpls:topology}. All other steps are performed in the same way as before.

\subsection{Runtime Modification of BGP Routers}
\label{subsec:bgp}

In the second scenario, we tested the ability of \systemname to apply changes to a BGP router in an already-deployed network. The scenario consists of a network with four routers, each one deployed in a different AS: \texttt{AS20} and \texttt{AS30} are two providers with peering; \texttt{AS200} is a customer of \texttt{AS30}; \texttt{AS100} is a customer of both \texttt{AS20} and \texttt{AS30}. We assume that \systemname is running in \texttt{AS100}, and it has already performed an initial translation from natural language to low-level router configurations, \ie it is aware of the topology and the running BGP configuration. To reach \texttt{AS200}, \texttt{AS100} uses the up-link towards \texttt{AS20} as the primary path, leaving the link towards \texttt{AS30} as a backup link
. Thus, the low-level configuration of \texttt{AS100} reduces the local preference attribute of the incoming BGP announcements sent by \texttt{AS30}. 
We use \systemname to switch the primary path of \texttt{AS100} from \texttt{AS20} to \texttt{AS30}.

To accomplish the desired configuration change using natural language, we ask ``Use \texttt{AS30} to reach \texttt{AS200}''. \systemname translates this new requirement into a waypoint specification. We observe that Step~2 is trivial as the system already knows the next hop towards \texttt{AS200}. In Step~3, the system generates the corresponding low-level commands in \texttt{vtysh} format. \systemname satisfies this requirement in different ways. In some cases, it removes the local preference modification on the \texttt{AS30} peering and applies it to\texttt{AS20}. In other cases, it defines a new route map that increases the local preference towards \texttt{AS30}. In some experiments, the generated configurations are semantically correct, but they contain some syntax errors
. A complete evaluation of GPT-4 ability to configure BGP remains as future work.
\section{Conclusion}
\label{sec:conclusion}

The rise of \glspl{LLM}, such as GPT-4, has provided a means to develop good quality code from AI-assisted tools. While multiple models have been proposed and fine-tuned for multiple programming languages, the same has not yet happened within the networking community. In this paper, we explore the opportunities to take advantage of \glspl{LLM} to improve network configuration. Our main takeaway is that \glspl{LLM} can dramatically simplify and automate complex network management tasks. We hope our work motivates more research on employing AI techniques on this matter.



\newpage
\bibliographystyle{abbrv}
\begin{small}
\bibliography{main}
\end{small}

\end{document}